\documentclass{iaus}
\usepackage{graphicx}

\newcommand{\kms}{\ifmmode {\rm km\,s}^{-1} \else km\,s$^{-1}$\fi}
\newcommand{\et}{\mbox{et~al.}\ }

\newcommand{\hb}{H$\beta$}
\newcommand{\civ}{C\,{\sc iv}}
\newcommand{\heii}{He\,{\sc ii}}
\newcommand{\mgii}{Mg\,{\sc ii}}
\newcommand{\lam}{$\lambda$}
\newcommand{\Mbh}{\ifmmode M_{\rm BH} \else $M_{\rm BH}$\fi}
\newcommand{\mbh}{\ifmmode M_{\rm BH} \else $M_{\rm BH}$\fi}
\newcommand{\Msol}{\ifmmode M_{\odot} \else $M_{\odot}$\fi}
\newcommand{\msol}{\ifmmode M_{\odot} \else $M_{\odot}$\fi}

\title[Quasar Mass Functions] 
{Quasar Mass Functions Across Cosmic Time}

\author[M. Vestergaard]   
{Marianne Vestergaard$^{1,2,3}$}

\affiliation{$^1$DARK Cosmology Centre, Copenhagen University, 
Juliane Maries Vej 30, 2100 Copenhagen \O, Denmark\\
Email: {\tt vester@dark-cosmology.dk}\\[\affilskip]
$^2$Steward Observatory, University of Arizona, 933 N. Cherry Avenue, Tucson, AZ 85721, USA\\[\affilskip]
$^3$Dept. of Physics and Astronomy, Tufts University, Robinson Hall, Medford MA 02155, USA  }

\pubyear{2010}
\volume{267}  
\pagerange{000--008}
\jname{Co-Evolution of Central Black Holes and Galaxies}
\editors{B.M.\ Peterson, R.S.\ Somerville, \& T.\ Storchi-Bergmann, eds.}

\begin{document}

\maketitle


\begin{abstract}
I present mass functions of actively accreting black holes detected in
different quasar surveys which in concert cover a wide range of cosmic
history. I briefly address what we learn from these mass
functions. I summarize the motivation for such a study and the methods by
which we determine black hole masses.

\end{abstract}

\firstsection 
\section{Motivation}

In cosmological studies there is a strong focus on the tight
relationships between the mass of the central black hole and that of
the stellar spheroidal component and its origin.  There are many other
reasons for our interest in the mass, demographics, and growth of
actively accreting black holes.  Briefly, theoretical models suggest
that the activity of black holes plays an important role for the
formation and evolution of mass structures in the universe, both at
the level of galaxies and galaxy clusters. In order to study and fully
understand the means by which black holes affect their small and large
scale environment and the physics underlying these processes, we need
to map the black hole mass and accretion rate
distributions. Establishing the black hole mass functions across
cosmic time is part of this crucial step.

\section{Black Hole Mass Estimates}

Peterson (these proceedings) gives a good
overview of how we determine masses of actively accreting black
holes. Therefore, I will only briefly outline the method we apply in
this work.

\subsection{Mass Scaling Relations Used Here}

The majority of our database consists of single-epoch spectra of 
active galaxies and quasars at large distances. Hence, we cannot 
directly utilize the variability properties of the sources to obtain a 
black hole mass based on the reverberation mapping (RM) technique. 
Instead we use the so-called ``mass scaling laws'' based on line 
width and continuum luminosity measurements from the single-epoch 
spectra (e.g., Vestergaard 2002). We use the results from RM that 
the size of the broad-line region $R$ scales with the nuclear 
continuum luminosity $L$ as $ R \propto L^{0.5}$ (Bentz \et 2006, 2009). 
Specifically, we use the mass scaling laws for \hb\ and \civ\ as 
published by Vestergaard \& Peterson (2006) and the scaling law for 
\mgii\ from Vestergaard \& Osmer (2009; hereafter VO2009). 

An important characteristic of these scaling relations is that they 
are calibrated to the same mass scale. While other \mgii\ relations 
exist in the literature, this is no other \mgii\ relation which 
shares the same mass scale as these or other \hb\ and \civ\ relations. 
For example, the \mgii\ relation of McLure \& Dunlop (2004) yields 
mass estimates that are up to a factor of 5 lower than the mass estimates 
of \hb\ and \civ\ for the same quasars (Dietrich \& Hamann 2004). The 
\hb\ and \civ\ relations used here are directly calibrated to the 
updated RM masses for the RM sample as published by Peterson \et (2004). 
The \mgii\ relation cannot be calibrated the same way since we only have 
one satisfactory measurement of the size of the \mgii\ emitting region 
to date (Metzroth \et 2006).  Therefore, we used high signal-to-noise 
SDSS spectra 
to inter-calibrate the \mgii\ scaling law to match the \hb\ and \civ\ 
relations of Vestergaard \& Peterson (2006).  All three scaling laws 
yield mass estimates that are consistent within the errors and which 
have absolute $1\sigma$ uncertainty of a factor of 3.5 to 4. The reader 
is referred to Vestergaard \& Peterson (2006) and VO2009 for further details.

We note that we have not taken the effects of radiation pressure into 
account in this work. The main reasons are that it is not entirely 
clear yet how important this effect is, if at all, and at present 
there is no prescription available for the \mgii\ and \civ\ lines. 
We will make corrections to the black hole mass functions for 
radiation pressure effects when this issue has been resolved.

\subsection{A Word of Caution}
At this meeting we have heard a few speakers mention that mass 
estimates based on the \civ\ emission line have serious issues and 
that the \civ\ line is not appropriate for mass estimates.  Since I use this method, it 
is important that I briefly address this issue. 

\begin{figure}[t]
 \vspace{-2.45 cm}
 \hspace{-1.3 cm}
 \includegraphics[width=2.2in]{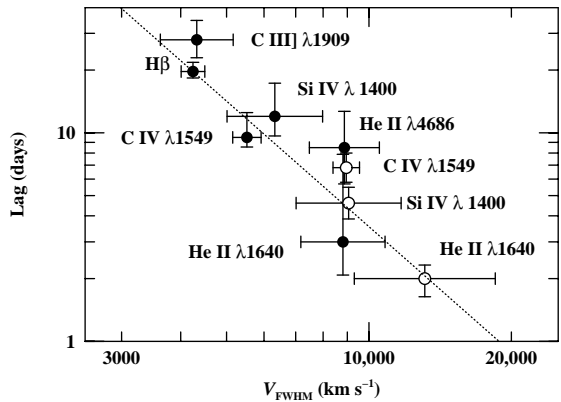} 
 \hspace{3.30 cm}
\begin{minipage}{2.0 in} 
 \vspace*{-5.9 cm}
\caption{Inverse relationship between response time delay (lag) and 
the velocity measured of the variable broad emission line gas for NGC 
5548 at two different epochs: 1989 and 1993. The dashed line shows the 
relationship for a virial mass of $6.8 \times 10^7$ \Msol.  
[Figure\ 1 of Peterson \& Wandel (1999). 
Figure is courtesy of Brad Peterson.]
   \label{fig1}}
\end{minipage}
\end{figure}

First, I would like to offer a word of caution: if the intent of a study 
is to compare mass estimates based on different lines, one should use 
mass scaling laws that are calibrated to the same mass scale. Otherwise 
the mass estimates are indeed expected to differ! All the works in the 
literature to date that compare mass estimates based on the \civ\ 
and \mgii\ emission lines do not use mass scaling laws that are 
on the same mass scale. Also, if two emission lines are present in your 
spectrum I strongly encourage you to use both emission lines to estimate 
the intrinsic black hole mass, since both lines contain important information. 
One should not discard one emission line because of being under the impression 
that it is less reliable. Peterson (these proceedings) shows that all 
the emission lines that have been monitored show that the dynamics of the 
line emitting gas is dominated by the gravity of the black hole.  This is 
the virial relation.  As the source luminosity changes, the line measurements 
(size $R$ and velocity $v$) slide up or down this relation. However, there 
is some scatter in the virial relation; the measurements do not all fall 
right on the relation (Figure~\ref{fig1}).  If you have only one measurement (or 
one line, e.g., \heii\,\lam{}4686 in Figure~\ref{fig1}) the mass estimate 
based thereon can be off from the intrinsic mass value; this offset we can 
only constrain statistically. 
Thus, the more lines and measurements we have the better constrained is 
the mass estimate.  The mass estimates I present here include multiple 
lines when they appear in the spectral window. I weight the individual 
mass estimates from each emission line according to their measurement 
uncertainties in computing the (final) black hole mass (see Vestergaard 
\et 2008 and VO2009). 

Second, it is not entirely clear that the issue with \civ\ is 
as bad as sometimes portrayed.  The amplitude of the deviations 
reported between \mgii\ and \civ\ mass estimates is for most 
studies of order 0.2$-$0.3 dex (e.g., Bachev et al. 2005; Baskin 
\& Laor 2005. See also Vestergaard \& Peterson 2006; Netzer 
\et 2007), well within the uncertainties associated with 
the mass scaling laws.  The most strongly deviating \mgii\ and 
\civ\ mass estimates are seen for samples with a large fraction 
of low signal-to-noise data (e.g., Shen \et 2008).  The high 
signal-to-noise SDSS data that we use to calibrate the \mgii\ 
scaling law do show a good correlation between the line widths 
of \mgii\ and \civ ; although some scatter is seen, the correlation 
between the mass measurements is somewhat stronger. It is worth keeping 
in mind that RM shows the \civ\ line gas to be virialized like 
the other emission lines.  Note, however, that the \civ\ scaling law is not 
useful for narrow-line Seyfert 1 galaxies (Vestergaard 2004b).

Our work on all three emission lines shows that none of them are 
perfect; systematic differences are in fact seen in their profiles 
and their line widths. Indeed, we do see stronger deviations between 
line widths and mass estimates based on \mgii\ and \civ\ than between 
\hb\ and \mgii . But at present the issue has not been investigated 
well enough to firmly question the validity of \civ\ as a mass 
estimator. For example, could the dependence of the \mgii\ profile 
width on the Eddington luminosity ratio (Onken \& Kollmeier 2008) 
play a role here?  We are in the process of investigating these issues 
regarding deviations between the mass scaling laws and will discuss 
our results in more detail in future work.

\section{Black Hole Mass Functions}

The differential quasar mass function is the space density of black 
holes per unit black hole mass as a function of redshift and black 
hole mass. We refer the interested reader to Vestergaard \et (2008) 
and VO2009 for further details on the determination of the mass 
functions that we present here.  

\subsection{Why Are the Quasar Luminosity Functions Not Enough?}

In the past, quasar luminosity distributions and luminosity functions 
have been used exclusively to constrain quasar and hence black hole 
evolution with cosmic time. So why do we even need the mass functions?  
It is worth keeping in mind that the luminosity is determined by the 
mass accretion rate.  The accreted mass $dM/dt$ is converted into 
radiation with a certain efficiency $\eta$, defined by $L = \eta c^2 dM/dt$. 
However, $L$ by itself does not tell us exactly how massive the central source 
is.  For example, there is no way of knowing if a certain source is a 
low-mass black hole with a very large mass accretion rate, or whether 
it is a massive black hole with moderate accretion rate. By estimating 
the black hole masses we now know that most of the black holes in the 
distant universe are, in fact, quite massive (e.g., Vestergaard 2004a;
Figure~1 of VO2009). 
If indeed the black holes we see as quasars were small but highly 
accreting we would expect to see narrow-lined quasars 
hovering right above the flux limits (dashed line in Figure\ 1 of VO2009) 
--- the opposite of what we observe. To our knowledge there is nothing 
preventing us from detecting quasars with broad line widths of 
$\sim$1000\,\kms\ to 2500\,\kms, yet we detect only very few of such 
sources among distant quasars. This is an important fact to keep in mind.  

Another issue regarding the sole use of the luminosity function is 
the unknown value of the radiative efficiency $\eta$ and how it 
may change with time, type of source, and/or the evolutionary stage 
of the source --- or, say, the spin of the black hole (e.g., Elvis \et 
2002). However, because $L$ is also a function of the black hole mass,
the anticipation is that by combining our measurements of the 
luminosity and the mass of the black hole we can break the degeneracy 
that currently exists in our use of the luminosity function alone 
(Wyithe \& Radmanabhan 2006) and perhaps even enable observational 
constraints to be placed on $\eta$.

\subsection{Quasar Samples and Data}
We present a summary of the mass functions of active black holes based 
on several large quasars samples recently published by Vestergaard \et 
(2008) and VO2009: 
the SDSS DR3 quasar sample, the Bright Quasar Survey (BQS), the Large 
Bright Quasar Survey (LBQS), and the SDSS Fall Equatorial Stripe 
redshift 4 sample. The reader is referred to these works for details 
on the samples and the data.  These samples are chosen because (a) a 
luminosity function has been established for each of them, (b) they 
each contain a large number of quasars, and (c) they complement each 
other well owing to the different selection criteria applied to the 
surveys in which they were found. Collectively, they cover a wide range 
of cosmic history, quasar properties, and a large part of the sky.  
For example, the SDSS survey is highly incomplete around redshifts of 
$z \sim 2$ to $z \sim 3$ due to the 
coincidence in color-space of quasars at those 
redshifts and the stellar locus. The LBQS of $\sim$1000 quasars 
complements well the SDSS since its quasars were selected based on SED 
shape in objective prism surveys. The implications are outlined later. 

\subsection{Sloan Digital Sky Survey DR3}
Richards \et (2006) present the luminosity function of a subset of 
about 15,000 DR3 quasars in a 1622 square degree area of the sky for 
which the selection function could be well established. They find a 
high-end slope of $\beta \approx -3.3$, and a steady decline in the 
amplitude from $z \sim  2$ to the local universe and from $z\sim 3$ 
to $z\sim5$.  For example, for a magnitude of 
$M_i = -27$\,mag, the luminosity function 
has dropped by a factor of a few hundred between $z \sim 2$ and 
$z\sim 0.5$. 
This shows that the space density of quasars that are this luminous 
has dropped dramatically in the equivalent time span.

\begin{figure}[tb]
\begin{center}
 \includegraphics[width=2.5in,angle=90]{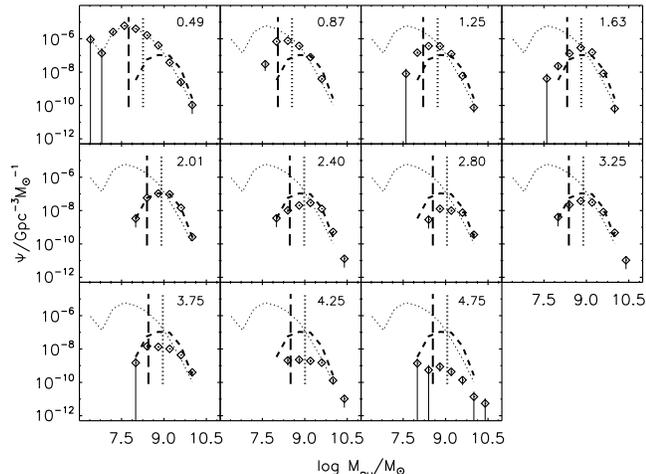}  
 \vspace*{-0.2 cm}
 \caption{Mass function of actively accreting black holes in the SDSS 
DR3 quasar sample as a function of mass for different redshift bins.  
The dotted and dashed curves show, for reference, the observed mass 
functions at redshift bins of 0.49 and 2.01, respectively. The vertical 
(dashed and dotted) lines mark the black hole mass of a fiducial quasar 
with a luminosity at the SDSS flux limit and emission-line widths of 
2000\,\kms\ and 3500\,\kms, respectively.
   \label{fig2}
}
\end{center}
\end{figure}

Figure~\ref{fig2} shows the mass function for this sample.
The characteristic turn-over toward low mass is mostly not real, but 
due to incompleteness.  This is mainly caused by the fact that we do 
not select quasars for our surveys based on black hole mass, but on
observed flux. For 
a given luminosity, a black hole can have a range of masses; in reality, 
a black hole of a given mass can have a range of luminosity because it 
may have a range of radiative efficiencies and/or mass accretion rates. 
The vertical lines denote the SDSS flux limits folded by, respectively, 
a line width of 2000\,\kms\ and 3500\,\kms, the median \hb\ line width 
in the local universe. They indicate that most of the turn-over is due 
to incompleteness. Quasars are seen to lie below these crude limits 
because the observed total luminosities 
(used for source selection) also contain contributions from stars in the 
host galaxy plus iron and Balmer continuum emission in addition to the 
nuclear power-law continuum luminosity used for the mass estimates. 

\begin{figure}
 \vspace*{-0.10cm}
\begin{center}
 \includegraphics[width=3.5in,angle=90]{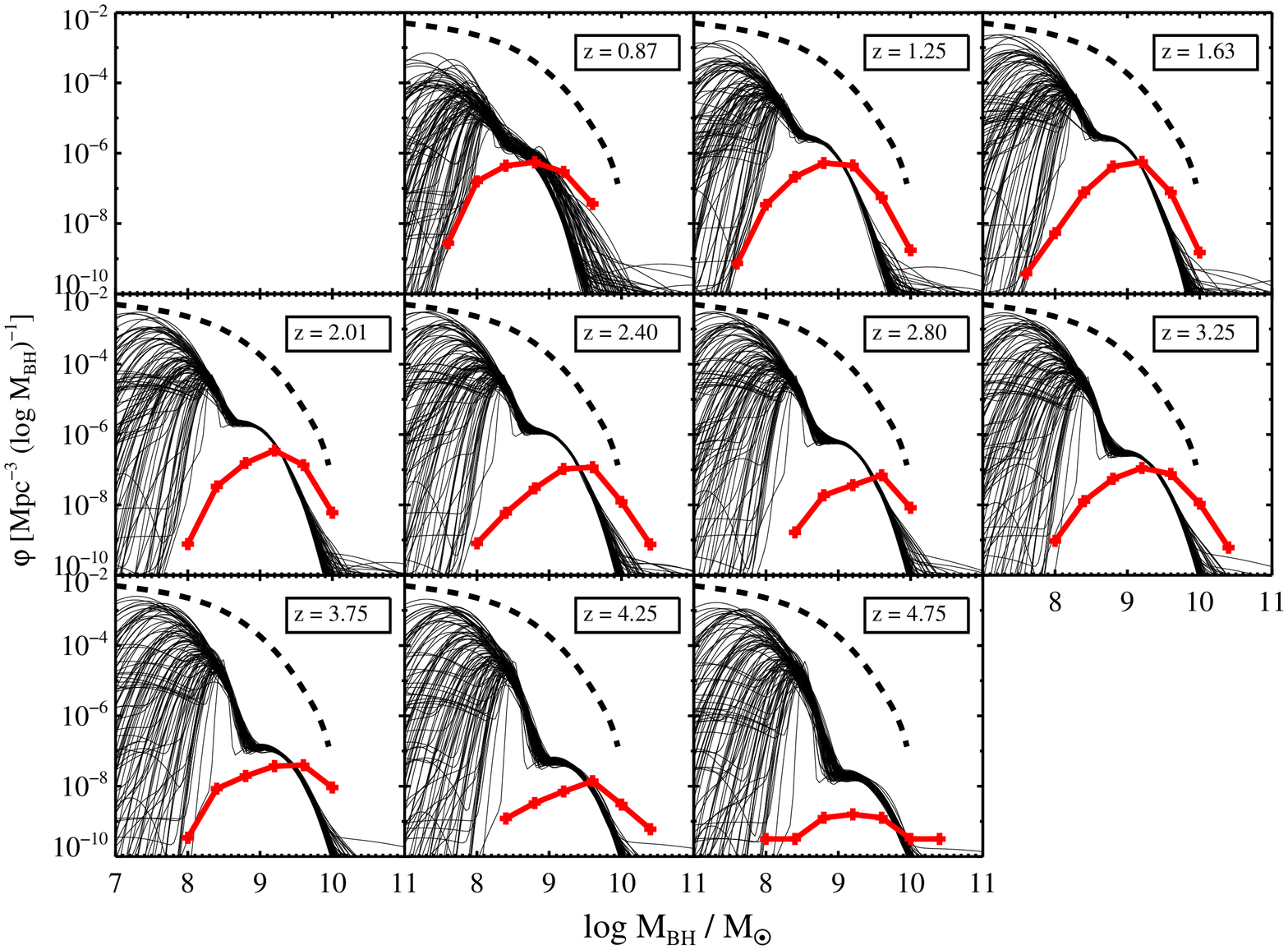}  
 \vspace*{-0.2 cm}
 \caption{ 
Random draws from the posterior probability density distribution of 
the underlying true black hole masses as established using Bayesian 
statistical methods that take into account the survey flux limits 
and uncertainties inherent in the mass estimates (light solid lines). 
The spread of the random draws illustrate the uncertainty in the mass 
functions.  The lower limit in the selection function is located at the 
indent between the two Gaussian curves; the mass functions are most 
trustworthy rightward thereof.  The binned mass functions from 
Figure~\ref{fig2} are shown as dots connected by solid lines.  
The dashed curve denotes the local mass function.
} \label{fig3}
\end{center}
\end{figure}

The high-mass end of the mass functions are nearly constant and 
those for black holes at $z \lesssim 3.5$ are all consistent
with a slope of $\beta = -3.3$, similar to the luminosity functions. However, 
in contrast, the amplitude of the mass functions only changes by a factor of 
a few.  The combination of the luminosity and mass functions thus show that 
the cumulative volume density of $L/M$ 
($\propto L/L_{\rm Edd}$) decreases toward 
low redshift. 

Because quasar surveys do not select sources based on black hole mass the 
binned mass functions shown in Figure~\ref{fig2} are statistically biased. 
We are currently using Bayesian statistical modeling to take the survey 
flux limits and the uncertainties in the mass scaling laws and on the spectral 
measurements into account (Kelly \et 2008, 2009a, 2009b, and
these proceedings). Our first (preliminary) 
results are in Figure~\ref{fig3} compared with the binned estimate of the mass 
functions.  It is clear that the intrinsic mass function has a steeper high-end 
slope and higher peak amplitude than evidenced by the binned mass function. It 
is too preliminary to say whether the mass functions can be described by a single 
power-law, double power-law, or a Schecther function --- that is, whether the 
bend right around the limit where the selection probability goes to zero 
value is real or not. The updated, final version of the mass function suggests
that it is not real; see the published journal article of Kelly \et 2009b).

\subsection{Maximum Black Hole Mass}

From our Bayesian analysis we constrain the maximum black hole mass of
the mass distribution with high probability to lie between $\sim$1.5
and 4 billion \Msol; such a black hole is most probably located at
a redshift $z \sim 3 $ (Kelly, these proceedings).

\begin{figure}[thb]
 \hspace{0.2 cm}
 \includegraphics[width=2.4in]{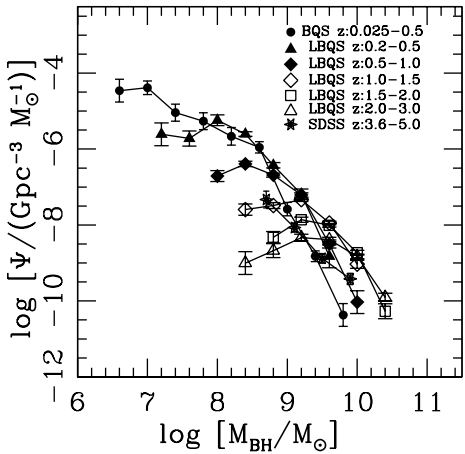}  
 \hspace{0.5 cm}
\begin{minipage}{2.2 in} 
 \vspace*{-7.0 cm}
 \caption{ 
Mass functions of the active black holes (as a function of mass 
for different redshift bins) based on the quasar samples of the 
Large Bright Quasar Survey (LBQS), the Bright Quasar Survey (BQS), 
and the SDSS sample of redshift 4 quasars in the Fall Equatorial 
Stripe (marked as SDSS and referred to as the ``SDSSz4'' sample in 
the text). These mass functions are presented by Vestergaard \&
Osmer (2009) [VO2009].
} \label{fig4}
\end{minipage}
\end{figure}

\subsection{The BQS, the LBQS, and the SDSS Fall Equatorial Stripe}

In Figure~\ref{fig4}, we compare the mass functions based on the BQS, 
the LBQS and the SDSS Fall Equatorial Stripe (hereafter, ``the 
SDSSz4 sample''). 
The LBQS sample has been divided into smaller redshift bins.   
Similar to what we saw for the SDSS DR3 sample, the low-mass 
turn-over is due to incompleteness. In addition, the high-mass 
end slopes are very similar. For the samples below $z\sim 3$ 
(BQS and LBQS) the high-end slopes are consistent with 
$\beta \approx -3.3$ as seen for the DR3 mass function. Above  
$z = 3.6$ (SDSSz4) the slope is shallower,  $\beta \approx -1.8$ (see 
VO2009). 
There is also a marked difference in amplitude: the space density 
of $z \approx 4$ active black holes is lower than at $z \approx 2$. 
This is discussed below.

\section{Evidence for Cosmic Downsizing}

\begin{figure}[t]
\begin{center}
 \includegraphics[width=1.7in]{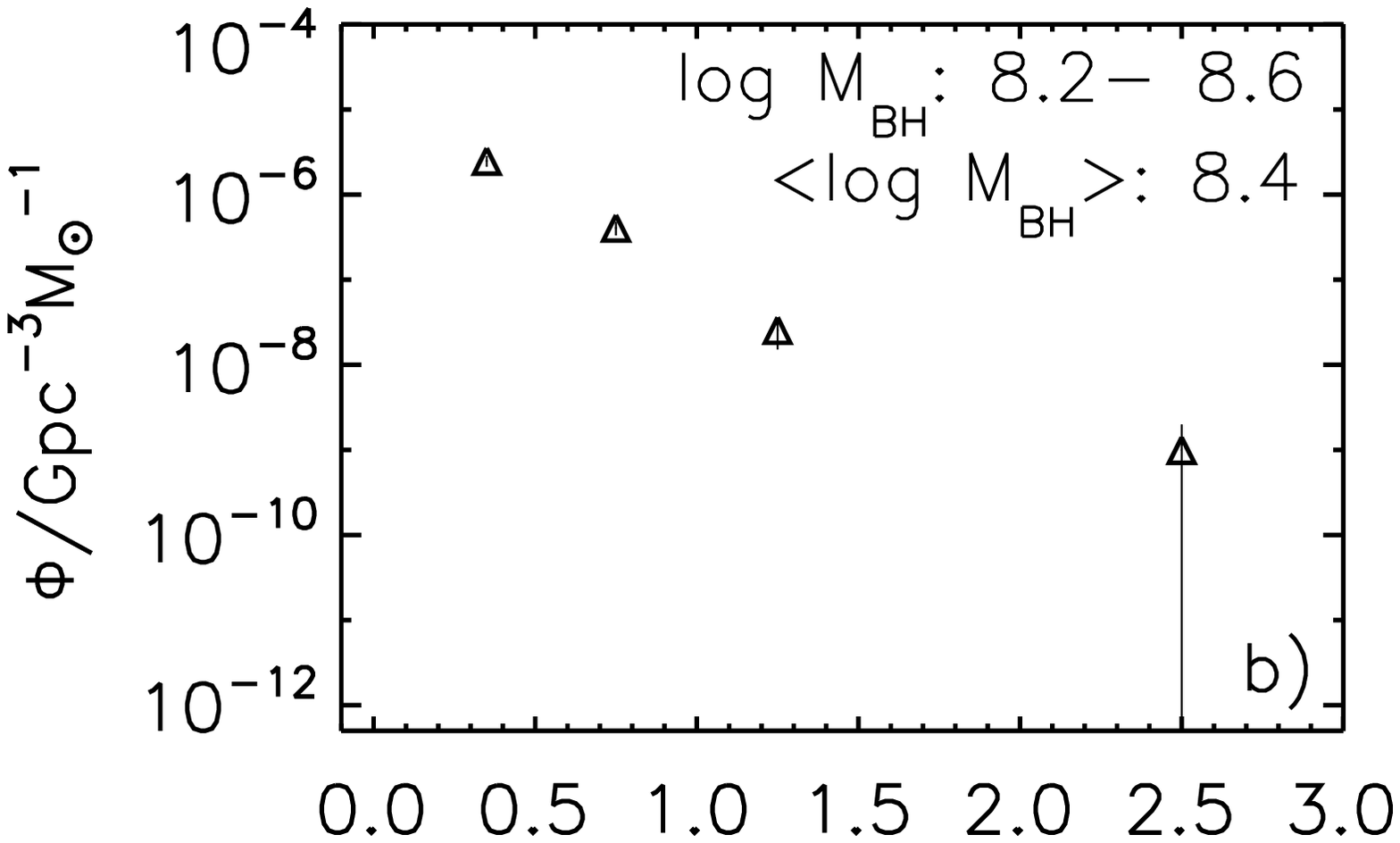}  
 \includegraphics[width=1.7in]{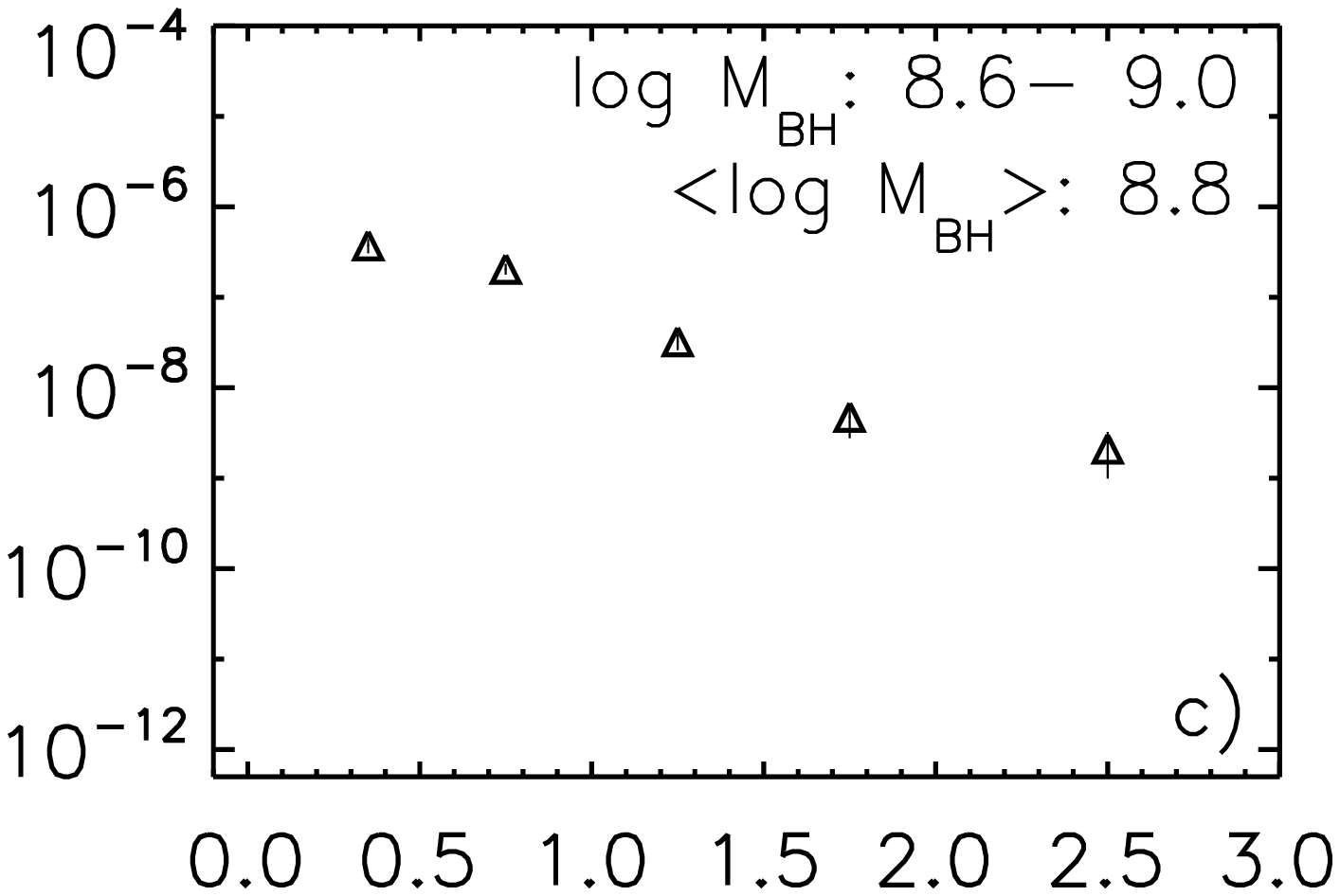}  
 \includegraphics[width=1.7in]{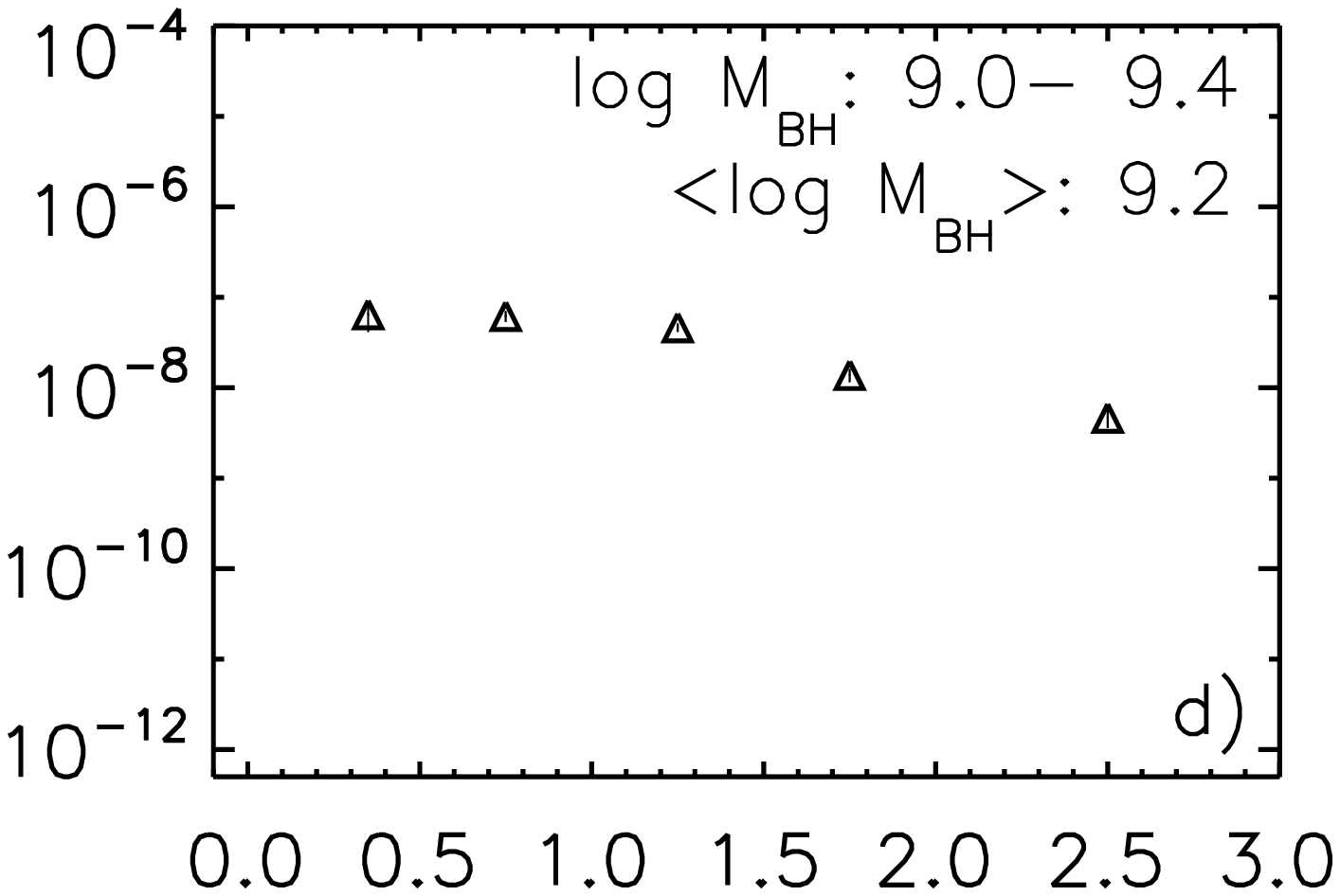}  
 \includegraphics[width=1.7in]{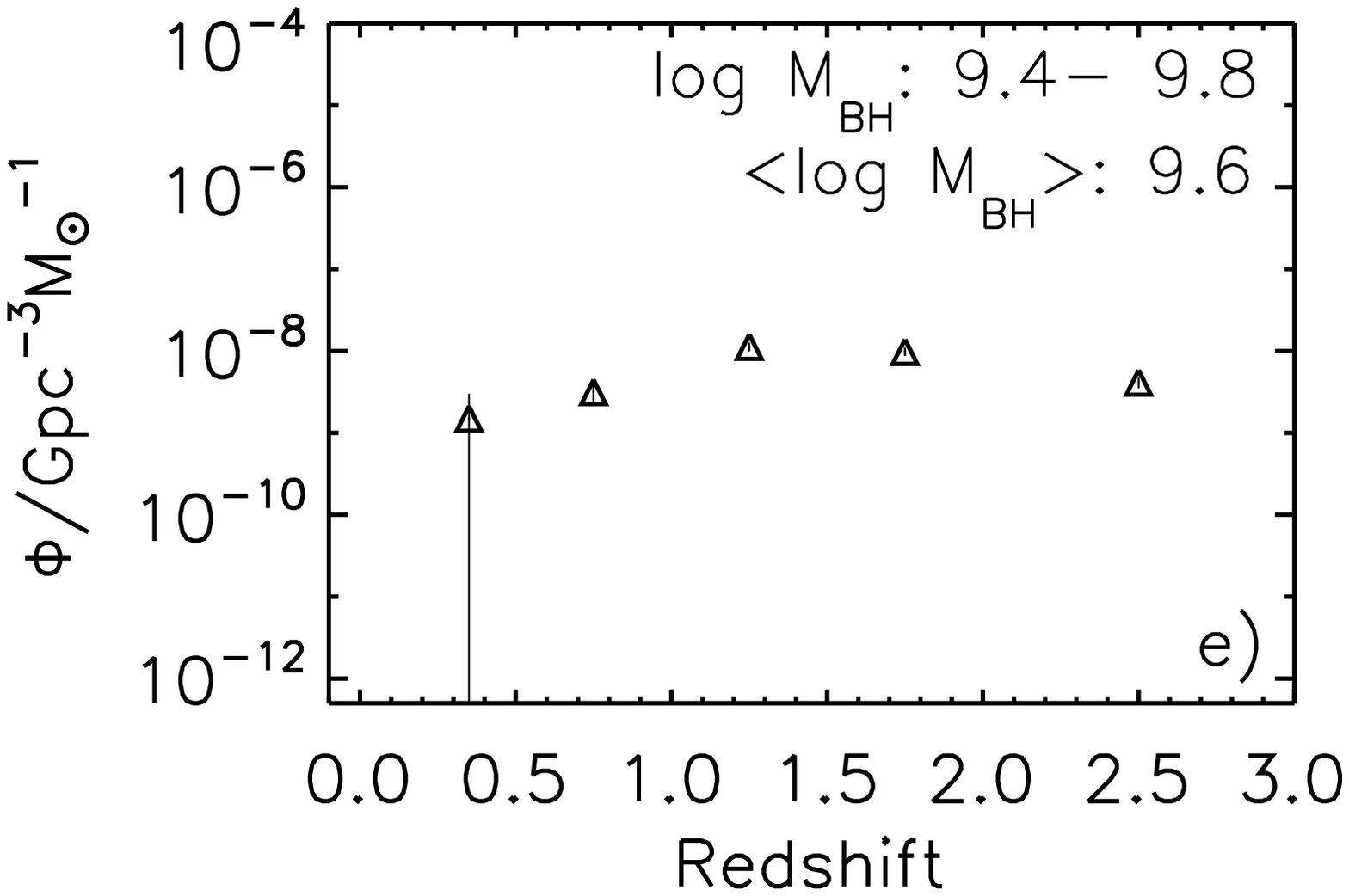}  
 \includegraphics[width=1.7in]{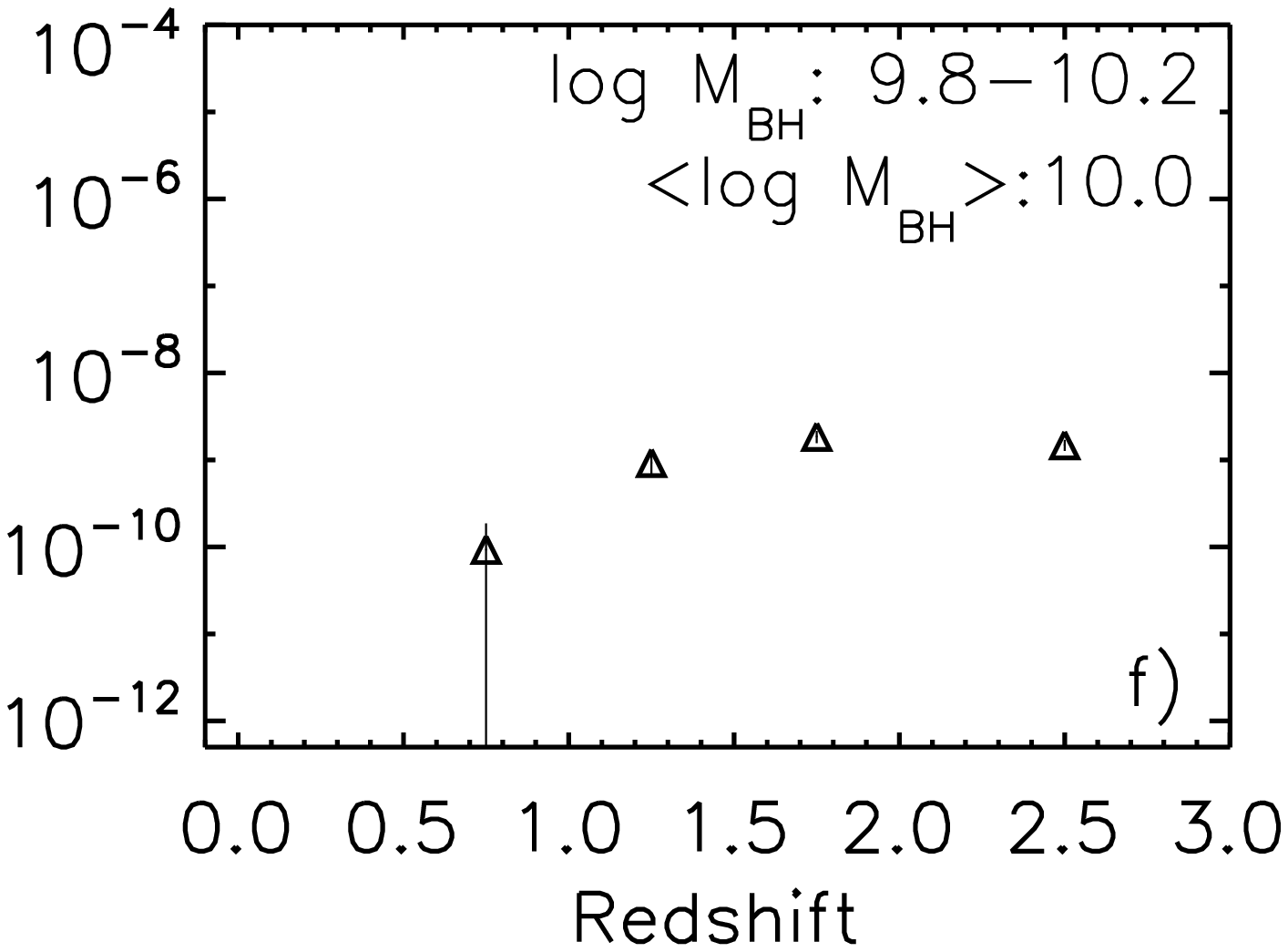}  
 \includegraphics[width=1.7in]{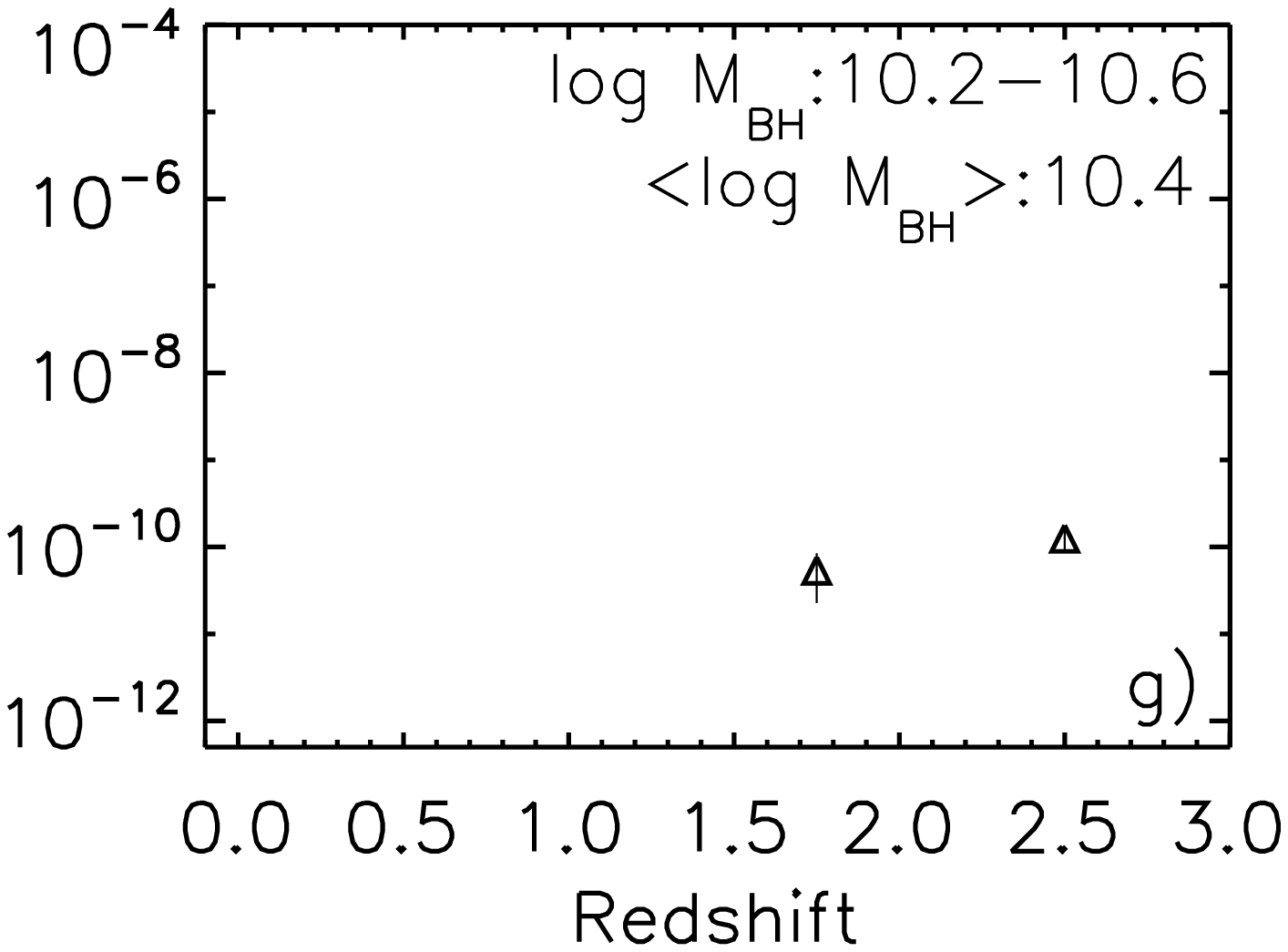}  
 \caption{ 
LBQS mass functions shown as a function of redshift for different mass bins. 
For a given mass bin, the redshift distribution shows at which epochs this 
active black hole is the most and least common. See Vestergaard
\& Osmer (2009) for the complete set of panels.
} \label{fig5}
\end{center}
\end{figure}

As noted earlier, the mass function is the space density of black holes 
as a function of both black hole mass and redshift.  Because the SDSS 
DR3 sample is highly incomplete around $z \approx$\,2 (and the selection 
function is prone to be more uncertain there) this sample does not add 
valuable information to the mass function as a function of redshift. 
But, the LBQS allows us to probe this very well.  In Figure~\ref{fig5} we 
show the space density distribution of black holes as a function of 
redshift for different mass bins, going from a mean black hole mass of  
$\sim 2.5 \times 10^{8}$ \msol\ (panel b) to a mean mass of $\sim 2.5 \times 10^{10}$ 
\msol\ (panel g). We see 
a steady progression of the maximum space 
density value from a low redshift to a higher 
redshift with increasing mass bin. This shows that the space density of 
active high-mass black holes was higher in the distant universe than it
is in the local one and the most active black holes now are of low mass. 
This is what we typically refer to as ``cosmic downsizing.'' 
This ``downsizing'' trend appears to extend to the local universe:
the mass function for local active black holes based on the Hamburg/ESO 
survey (Schulze \& Wisotzki 2008; their Figure 4) shows a higher amplitude 
for $\mbh < 10^8$\,\Msol{} than those based on LBQS and SDSS DR3 at 
$0.2 \lesssim z \lesssim 0.5$
while the high-mass end slope 
($\beta \approx -3.3$) is consistent
with those of the LBQS and SDSS DR3 mass functions.

\begin{figure}[h]
\begin{center}
 \includegraphics[width=2.4in]{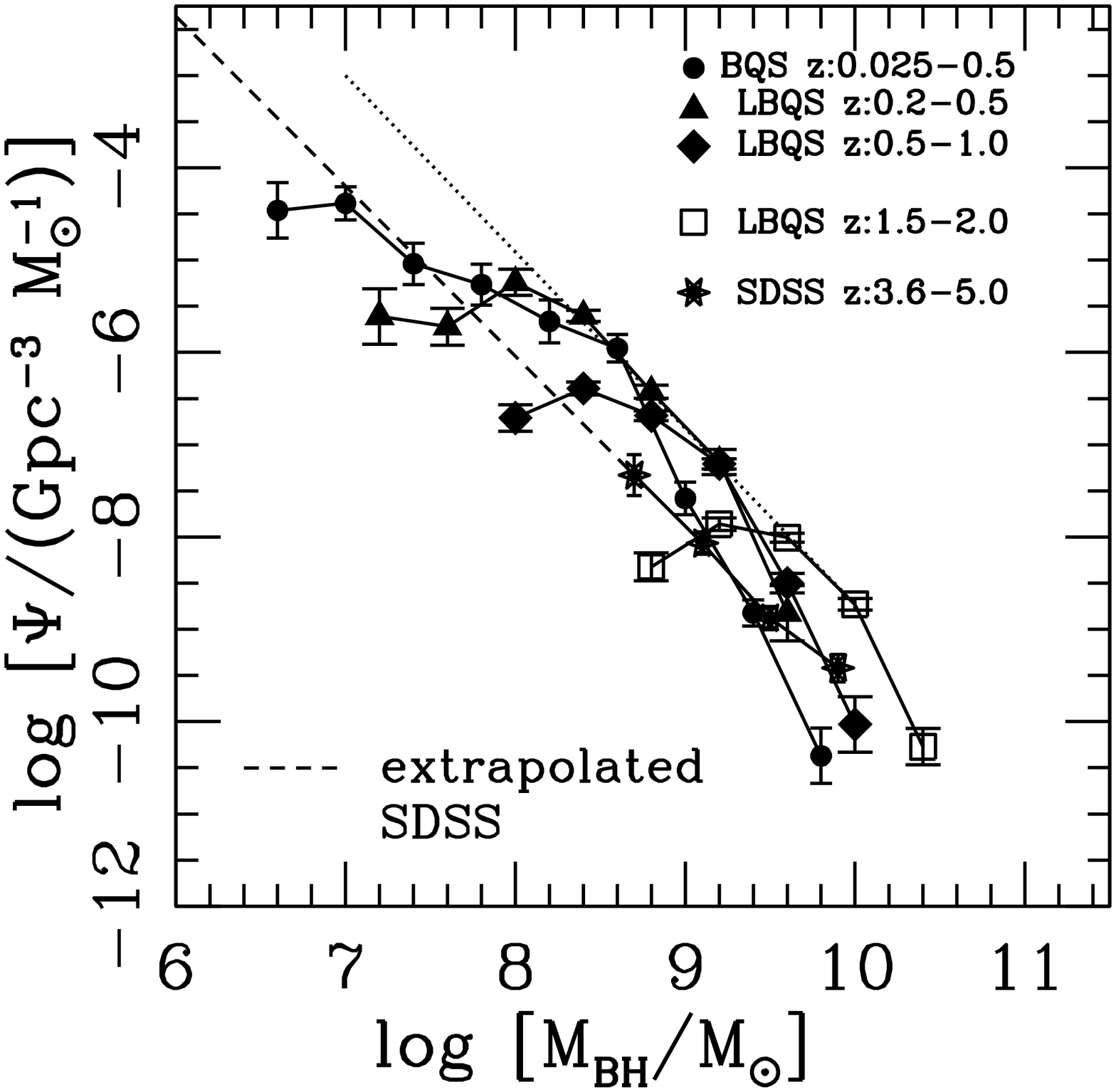}  
 \includegraphics[width=2.4in]{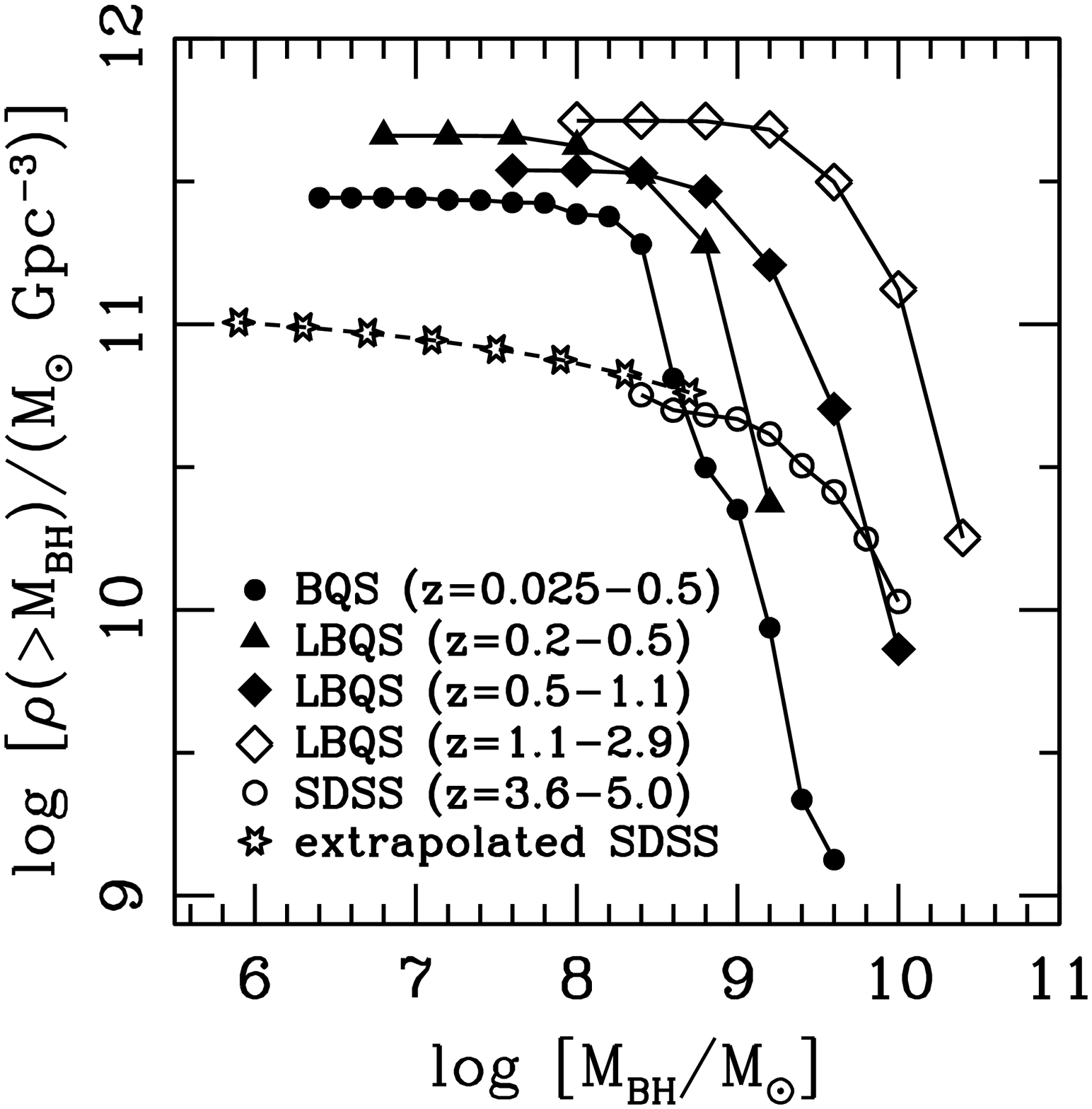}  
 \caption{ 
Mass functions (left panel) and cumulative mass density (right panel) for 
the BQS, LBQS, and SDSSz4 samples. In these diagrams the SDSSz4 mass 
function is extrapolated to very low masses and the implications for the
cumulative mass density (open stars) is tested.
} \label{fig6}
\end{center}
\end{figure}

\section{Evidence for Fast Build-up of the Black Hole Population at 
\boldmath{$z \sim 4$}}

We mentioned earlier that the mass function for the SDSSz4 sample 
has a distinctly different slope and amplitude compared to the 
LBQS and BQS mass functions at lower redshift (e.g., Figure~\ref{fig6},
left panel). We also see a 
distinctly lower amplitude of the cumulative mass density 
(Figure~\ref{fig6}, right panel) of this sample compared to the LBQS 
at $z \sim 2$ to $z \sim 3$. Could this be caused by our inability to detect 
the low-mass black holes (owing to the SDSS flux limits)?  As a 
crude test thereof we extrapolated the SDSSz4 mass function to very 
low masses (dashed line in left panel) as a case study for the event 
that we were able to detect all black hole mass values.  The result 
of integrating over the mass function and its extension is shown in 
the right panel as a dashed/asterics-dotted curve. This shows that 
our inability to detect black holes below a mass of $\sim 3 \times 
10^8$\, \msol\ does not explain the factor 10 lower cumulative mass 
density between $z \sim 4$ to $z \sim 2$ as seen in the right panel. 
In order for the cumulative
mass density at $z \sim 4$ to approach that at lower redshift, the 
distribution of low-mass black holes has to be somewhat steeper than 
that indicated by the dashed line in the left panel.  While future 
work will have to test such a scenario, we do not consider this very 
likely. Rather, the differences in mass functions and cumulative mass 
density between $z \sim 4$ and $z \sim 2$ suggest that a significant 
mass growth takes place during this time span.

The low-mass extension of the SDSSz4 mass function is parallel to 
the slope (dotted line, left panel) that the LBQS mass functions 
appear to move along with redshift. This amplitude difference 
between the dotted and dashed lines shows that on average the 
black hole space density needs to increase by about a factor of 
17 from $z \sim 4$ to $z \sim  2$.   
We note, tongue in cheek, that while the quasar luminosity functions 
also show a strong decline in space density of luminous quasars with 
increasing redshift,  this decline can just as well be explained by a 
decline in the number density of active black holes as a decline in 
their luminosities with redshift.  The mass functions help confirm 
that a space density decline and a mass density decline (i.e., 
the typical mass of the black holes declines) is taking place. 

\begin{acknowledgements}
I thank my collaborators for contributions that 
made this work possible: X. Fan (Arizona), B. Kelly (CfA), K. Denney, 
C. Grier, P. Osmer, B. Peterson (Ohio State),  
C. Tremonti (Wisconsin--Madison),  
M. Bentz (UC Irvine), H. Netzer (Wise Obs. Israel).  I acknowledge support 
from HST grants HST-GO-10417 (PI Fan), HST-GO-10833 (PI Peterson), and 
HST-AR-10691 from NASA through the Space Telescope Science Institute, 
which is operated by the Association of Universities for Research in 
Astronomy, Inc., under NASA contract NAS5-26555.
\end{acknowledgements}

{}

\end{document}